\documentclass[12pt,manuscript,draft]{aastex}

\input psfig.sty

\shorttitle{Search for Triaxiality}
\shortauthors{Larsen et al.}

\begin{document}

\title{Mapping the Asymmetric Thick Disk I.  A Search for Triaxiality}

\author{Jeffrey A. Larsen}
\affil{Physics Department, United States Naval Academy,
    Annapolis, MD 21402}
\email{larsen@usna.edu}

\author{Juan E. Cabanela}
\affil{Department of Physics and Astronomy, Minnesota State University Moorhead, Moorhead MN, 56563}
\email{cabanela@mnstate.edu}

\author{Roberta M. Humphreys}
\affil{Astronomy Department, University of Minnesota, Minneapolis MN, 55455}
\email{roberta@umn.edu}

\and

\author{Aaron P. Haviland}
\affil{Physics Department, United States Naval Academy,
    Annapolis, MD 21402}

\begin{abstract}

A significant asymmetry in the distribution of faint
blue stars in the inner Galaxy, Quadrant 1 ({\it l} $ = 20\arcdeg - 45\arcdeg$)
compared to Quadrant 4 was first reported by \cite{lar96}.
Parker et al (2003, 2004) greatly expanded the survey to determine its 
spatial extent and shape and the kinematics of the affected stars. 
This excess in the star counts  was subsequently
confirmed by \cite{jur08} using SDSS data. Possible explanations for
the asymmetry include a merger remnant, a triaxial Thick Disk, and a
possible interaction with the bar in the Disk. In this paper we describe
our program of wide field photometry to
map the asymmetry to fainter magnitudes and therefore larger distances.
To search for the signature of triaxiality, we extended our survey to 
higher Galactic  longitudes. We find no evidence for an excess of faint blue stars at
{\it l} $\ge 55\arcdeg$ including the faintest magnitude interval. The
asymmetry and star count excess in Quadrant 1 is thus not due to a triaxial
Thick Disk.

\end{abstract}


\section{Introduction: The Asymmetric Thick Disk}

Studies of both stars and gas are revealing significant structure and
asymmetries in their motions and spatial distributions including the bar of
stars and gas in the Galactic bulge (\citealt{1991ApJ...379..631B},
\citealt{1994ApJ...429L..73S}), the evidence from infrared surveys for a larger stellar bar in the inner
disk (\citealt{1992ApJ...384...81W}, \citealt{1997MNRAS.292L..15L},                
\citealt{2005ApJ...630L.149B}),
the discovery of the Sagittarius
dwarf
\citep{1994Natur.370..194I,1995MNRAS.275..591I}, and a significant asymmetry of unknown 
origin in the distribution of faint blue stars in Quadrant 1 (Q1) of the inner Galaxy 
\citep{lar96}. Each of these observations provides a significant clue
to the history of the Milky Way. When combined with the growing evidence 
for Galactic mergers in addition to the Sagittarius dwarf, i.e.
 the Monoceros stream (\citealt{2002ApJ...569..245N}, 
\citealt{2003MNRAS.340L..21I}), the Canis Major merger remnant 
(\citealt{2004MNRAS.355L..33M}),
the Virgo  stream (\citealt{2001ApJ...554L..33V}, 
\citealt{2005ApJ...633..205M}), we now realize that the
structure and evolution of our Galaxy have been significantly altered by 
mergers
with other systems.  Indeed the population of the Galactic Halo
and possibly the Thick Disk as well, may be dominated by mergers with 
smaller systems.

The  asymmetry in  Q1,  $l = 20\arcdeg - 45\arcdeg$,  
first recognized by \cite{lar96} was based on a comparison with complementary
fields in the fourth  Quadrant (Q4) using  star counts from the
Minnesota Automated Plate Scanner Catalog of the
POSS I \citep{cab03}\footnote{The MAPS database is online at:  
http://aps.umn.edu}. 
To map the extent and shape of the asymmetric distribution and further 
identify the contributing stellar population, \cite{par03}  extended 
the search to 40 contiguous fields from the digitized POSS I on each side 
of the Sun-Center line plus the same number of fields below the plane in Q1. 
They examined the star count ratios for paired fields in three color ranges: 
blue, intermediate and red.  Over 6 million stars were used in the star count 
analysis covering almost 2000 square degrees on the sky.
They found a 25\% excess in the number of probable Thick Disk stars in Q1, $l \approx 20$ to $60\arcdeg$ and $20\arcdeg$ to $40\arcdeg$ above and below the plane compared to the complementary fields ($l \approx 340$ to $300\arcdeg$) in Q4. While the region of the asymmetry is somewhat irregular in shape, 
it is also fairly uniform and covers several hundred square degrees. It is therefore a 
major substructure in the Galaxy due to more than small scale clumpiness. Assuming that these are 
primarily main sequence thick disk stars, with a typical magnitude completeness limit of $V \approx 18$ mag, they are 1 - 2 kpc from the Sun and about 0.5 to 1.5 kpc above (and below) the plane.

\cite{par04} also found an associated kinematic signature. Using velocities from spectra for more than 
700 stars, they not only found an asymmetric distribution in the v$_{LSR}$ velocities, but the Thick Disk stars in Q1 have a much slower effective rotation rate $\omega$ 
compared to the corresponding Q4 stars. A solution for the radial and tangential components of the v$_{LSR}$ velocities reveals a significant lag of 80 to 90 km s$^{-1}$ in the direction of Galactic 
rotation for the Thick Disk stars in Q1.

Three possible explanations for the asymmetry are the
fossil remnant of a merger, a triaxial Thick Disk or inner Halo,  and 
interaction of the Thick Disk/inner Halo stars with the bar in
the Disk. Given the  lack of any spatial overlap with the path of the
Sagittarius dwarf through the Halo \citep{2003MNRAS.340L..21I}, and the predicted 
path of the Canis Major dwarf \citep{2004MNRAS.355L..33M}, its   association  with either 
feature is unlikely. Furthermore its spatial extent and apparent symmetry relative to the plane also
do not automatically support a merger interpretation. Our  line of sight to the asymmetry is interestingly in the 
same direction as the stellar bar in the Disk \citep{1992ApJ...384...81W,2005ApJ...630L.149B}, and the near end of the Galactic bar \citep{1997MNRAS.292L..15L, 2000MNRAS.317L..45H}, but  the Disk bar is approximately 5 kpc from the Sun in 
this direction. Thus the stars showing the excess are between the Sun and 
the bar, not directly above it. However the maximum extent of the  star count excess 
along our line of sight is not known. 

Recently \citet{lar08} showed the stellar density distribution from the MAPS scans for the 
\citet{par03} fields in Q1 and Q4 above the plane, demonstrating  that the excess was in Q1 and was
not due to a ring above the plane \citep{jur08}. Larsen et al named the asymmetry feature the 
Hercules Thick Disk Cloud in recognition of the  direction  on the sky where the star count 
excess is maximum. Thus,  interpretation of the Hercules Thick Disk Cloud is not clear-cut.  
While it might well be a fossil merger remnant, the star count excess is  also 
consistent with  a triaxial Thick Disk or inner Halo as well as a gravitational 
interaction with the stellar bar especially given the corresponding asymmetry in the kinematics. 
The distinction between a triaxial Thick Disk and interaction with 
a Disk bar may be difficult to discern. Indeed, it is unclear which 
may have formed first, and one or both may be the result of mergers.

The release of the SDSS Data Release 5 (DR5) photometry in 
the direction of the observed asymmetry in 
Q1 led to the discovery of a feature at much fainter magnitudes, the distant Hercules-Aquila 
cloud \citep{bel07} and also confirmed the nearer star count excess in the 
inner Galaxy \citep{jur08}.  The SDSS survey however 
is not well designed for a comprehensive survey  of the Thick Disk inside the 
Solar orbit. It extends below $b = 30\arcdeg$ in only a few directions in Q1 
and has only limited coverage in Q4.  It does not have the leverage in 
Galactic longitude needed to discriminate among the  possible causes of the Hercules Thick Disk Cloud.

To further explore the possible origin of the observed spatial and kinematic 
asymmetry, we have obtained multicolor CCD images to extend the star 
counts to fainter magnitudes to map the spatial extent of the asymmetry 
along the line of sight as a function of Galactic longitude and latitude. 
For example, if the Thick Disk is triaxial we would expect to observe 
the star count excess out to greater longitudes, but 
the star count excess described by \citet{par03} appears to terminate 
near {\it l} $\sim$ 55$\arcdeg$. However, if it is triaxial in Q1 with 
its major axis above
the bar, the stars  would be further away at the higher longitudes.  
By extending the star counts to fainter magnitudes, corresponding 
to greater distances, 
we can search for the asymmetry at higher longitudes, 
from {\it l} of 50$\arcdeg$ to 75$\arcdeg$.

In this first paper we describe our observing program and present results on 
 a search for triaxiality of the inner Halo and Thick Disk.  Our second paper will discuss the star counts across the full range of longitudes in Q1 and Q4 
 containing the asymmetry feature and a third paper in the series will analyze 
 the kinematics of the associated stars. In the next section we describe the CCD observing program and data reduction. Determination of the star count ratios 
 and their errors for the fields at the higher
 longitudes are discussed in \S 3. Our  results  do not support a 
 triaxial interpretation of the asymmetry feature and are summarized in \S 4.

\section{Observations and Data Reduction}

\begin{figure*}
  \begin{center}
    \leavevmode
      \psfig{file=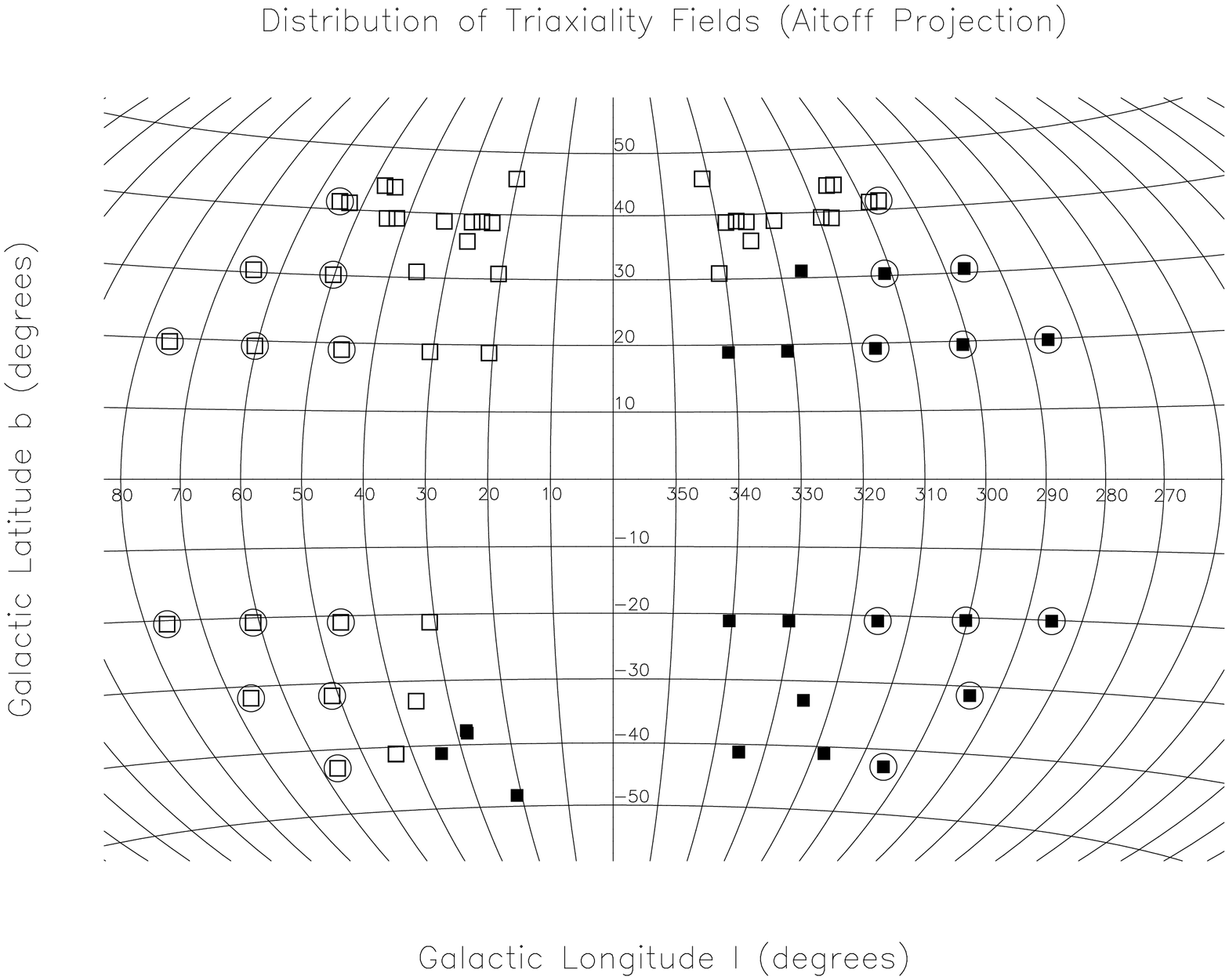,width=\textwidth,angle=-90}
        \caption{Distribution on the sky of  project fields.
Fields observed with 90Prime are 
plotted as open squared and fields observed with Y4KCam are represented 
as filled squares.  Fields used for our first result on triaxiality are circled.
The size of each icon is roughly twice as large
as the actual sky coverage of the corresponding field. }
     \label{fig1}
  \end{center}
\end{figure*}

To map the asymmetry feature both above and below the Galactic plane we 
identified 67 fields ranging in longitude from {\it l} = 20$\arcdeg$ to
75$\arcdeg$, and {\it l} = 340$\arcdeg$ to 285$\arcdeg$ and latitude 
{\it b} = $\pm$20$\arcdeg$ to $\pm$45$\arcdeg$.  The total sky coverage of the program is 47.5 square degrees.  The  distribution of our 
program  fields on the sky is shown in Figure~\ref{fig1} and the  information 
for each field is given in Table~\ref{tbl-0} with their Galactic and equatorial
centers, instrument, field size and date of observation \footnote{Each field is identified according to its Galactic longitude and latitude with an {\it H} to distinguish it from similar fields based on APS data}.   In this paper  
we will be discussing a subset of the fields relevant to the  
triaxiality question.  These fields are circled on Figure~\ref{fig1} and are listed in Table~\ref{tbl-3}.  Our program required observations in both 
hemispheres.

\subsection{Northern Hemisphere - Steward Observatory Bok 2.3 Meter}

We used the Steward Observatory 2.3 meter Bok telescope with the 1 square degree 90Prime camera in 2006 May, 2007 September and 2008 May.  This mosaic camera has 4 blue sensitive 4096x2096 CCDs.  Counting cosmetic defects and the plate scale of 0.45\arcsec~per pixel we had an effective field of view of approximately 1.02 square degrees.  The CCDs are arranged in a square with intra-CCD gaps of about 10 arcminutes.  We used a Johnson U,B,V + Cousins-Kron R filter set for the observations.  This field of view was ideal for the project as it allowed each of the project fields to be observed in a single pointing.  

Program fields were observed twice in each filter with exposure times of 
600 seconds in U, 480 seconds in B, and 90 seconds each in  V and R.
Exposure times were chosen to reach a limiting magnitude of 20th in U and 22nd in B, V and R.
These exposures were of sufficient length to require autoguiding.
The filters were not parfocal but focus was digital and corrections were 
easily applied.

We reduced the data with the IRAF \it{MSCRED }\rm mosaic image reduction
package \citep{1998ASPC..145...53V}.  The data were corrected for cross talk,
bias, and dark current.  Flatfielding was done using night-sky flats 
from program fields supplemented by a series of random sky exposure with
equivalent exposure times.  CCD defects, bad pixels and satellite trails 
were removed.

Astrometry was done with Larsen's MOSAF (Mosaic Astrometry Finder)
program written for the Spacewatch 0.9 meter telescope \citep{2007AJ....133.1247L} which 
uses a 13 term polynomial for a plate model, the CFITSIO library \citep{1999ASPC..172..487P} 
and WCSLib from WCSTools \citep{1999ASPC..172..498M}.  The astrometric reference catalog was USNO-B1.0 \citep{2003AJ....125..984M}, and the residuals were approximately 0.3\arcsec~ in all frames.

Photometric reductions used both Landolt Standards \citep{1992AJ....104..340L} and a secondary set of faint standards from \cite{1998ApJS..119..189O}, and    followed the method of \cite{1964aste.book..178H}.  The formal 
zero point errors were approximately 0.03 magnitudes per filter with 
the photometric RMS of 0.05 magnitudes per band.  An example of the mean 
atmospheric extinction and color terms can be found in Table~\ref{tbl-5};
seeing  varied from 1.0\arcsec~ to 2.8\arcsec~ during the 2006 May observations.  

\subsection{Southern Hemisphere - CTIO SMARTS 1.0 Meter}

Fields which could not be reached  with the Bok 2.3 
meter telescope were observed using the Y4KCam at the 1.0 meter SMARTS consortium 
telescope at CTIO in 2006 April and 2008 October.  This camera is a 4096x4096 pixel single 
CCD with a plate scale of 0.289" per pixel and a field of view of about 0.108 square degrees.  
Consequently, 9 subfields were observed to yield an effective area of $\approx$ 0.95 square
degrees with overlap.  The  Johnson U,B,V + Cousins-Kron R filter set was used for these observations.

Program fields were observed once in each filter.  During  2006 April, the 
exposure times were 
300s in U, 150s in B, and 60s  in the  V and R bands.
This was sufficient to reach 19th magnitude in B. V and R and 18th magnitude in U. 
In 2008 October, the exposure times were increased to 400s in U, 485s in B,
160s in V and 190s in R to reach fainter magnitudes.
These exposures were of sufficient length to require autoguiding.
The filters were not parfocal but could be compensated by a focus reduction.

 Our initial processing of the frames was  based on the IRAF scripts for the Y4KCam provided by 
 Massey\footnote{Available at: http://www.lowell.edu/users/massey/obins/y4kcamred.html}. 
 Flat fielding used both dome and twilight flats.
Astrometry was done using WCSTools and the USNO-B1.0 catalog with results refined and checked against MOSAF.  Seeing was moderate for the 2006 April observations,  with FWHM from 1.5\arcsec~ to 3.0\arcsec~.

Photometric reductions used both Landolt Standards \citep{1992AJ....104..340L} and a secondary set of faint standards from \cite{1998ApJS..119..189O}.  Photometric magnitudes were then determined using the same method as the 90Prime data 
to assure consistency. Examples of the mean atmospheric extinction and color terms are in Table~\ref{tbl-6}.

\subsection{Catalog Creation}

Source detection, image parametrization and catalog creation used Source Extractor  \citep{1996A&AS..117..393B}.  After several tests to  verify that we were detecting objects close to the image limits, the default size for the Gaussian convolution mask was used and magnitudes were determined using the MAG\_AUTO feature which uses an adaptive aperture photometry routine.  Aperture photometry was performed on the detected objects.  Since most are stars, they were not subject to the same faint-end systematics as galaxies \citep{2002ApJ...571...56B}.  We also discarded all objects detected within 4\arcsec~ of 
the edge of the frame as  potentially truncated.

A custom  pipeline was written to apply the WCS astrometry solution and photometric calibration to  each  object in the catalogs for each field. We then used 
the WCS positions to match the objects observed in the different filters.  To
be included in the final catalog, each object had to be measured in at least 
two filters to provide color information.  Problems with the star-galaxy 
classifier in Source Extractor are discussed in the next section.

The final catalogs with magnitudes, positions, colors and object classifications will be available on-line with the second paper.

\subsection{Star Galaxy Separation}

\begin{figure*}
  \begin{center}
    \leavevmode
      \psfig{file=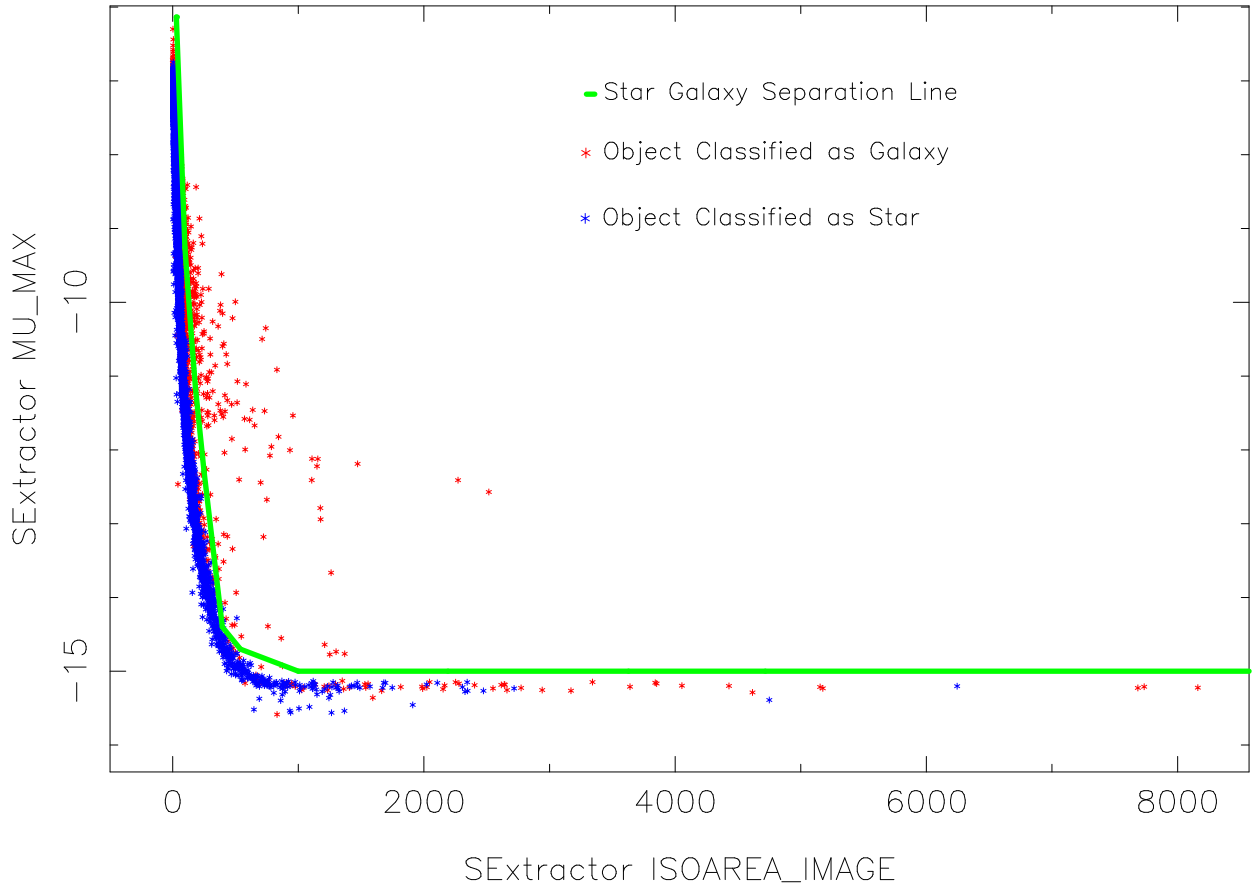,width=\textwidth,angle=-90}
        \caption{Star/galaxy separation with MU\_MAX vs. ISOIMAGE\_AREA. The stellar locus is well defined.  The plot represents all points on amplifier "A" of 90Prime during a 120 second R band exposure  on the night of May 26, 2006.  The discrimination line is plotted in green.  Objects which are classified as a galaxy when all three discriminators are applied are plotted in red while objects which remain on the stellar locus are plotted in blue.  }
     \label{fig2}
  \end{center}
\end{figure*}

\begin{figure*}
  \begin{center}
    \leavevmode
      \psfig{file=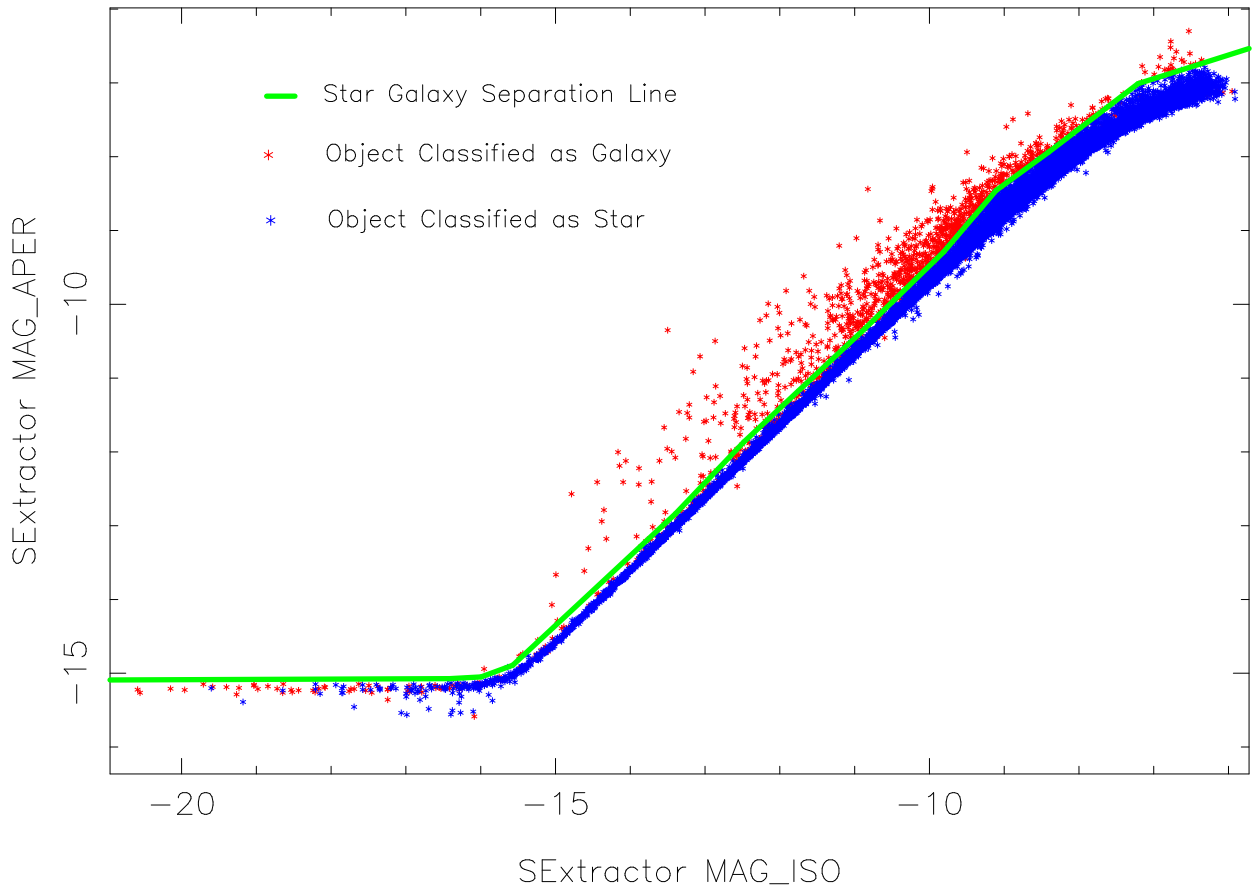,width=\textwidth,angle=-90}
        \caption{Star/galaxy separation with  MAG\_ISO versus MAG\_APER.  The plot represents all points on amplifier "A" of 90Prime during a 120 second R band exposure  on the night of May 26, 2006.  The discrimination line is plotted in green.  Objects which are classified as a galaxy when all three discriminators are applied are plotted in red while objects which remain on the stellar locus are plotted in blue.  }

     \label{fig3}
  \end{center}
\end{figure*}

\begin{figure*}
  \begin{center}
    \leavevmode
      \psfig{file=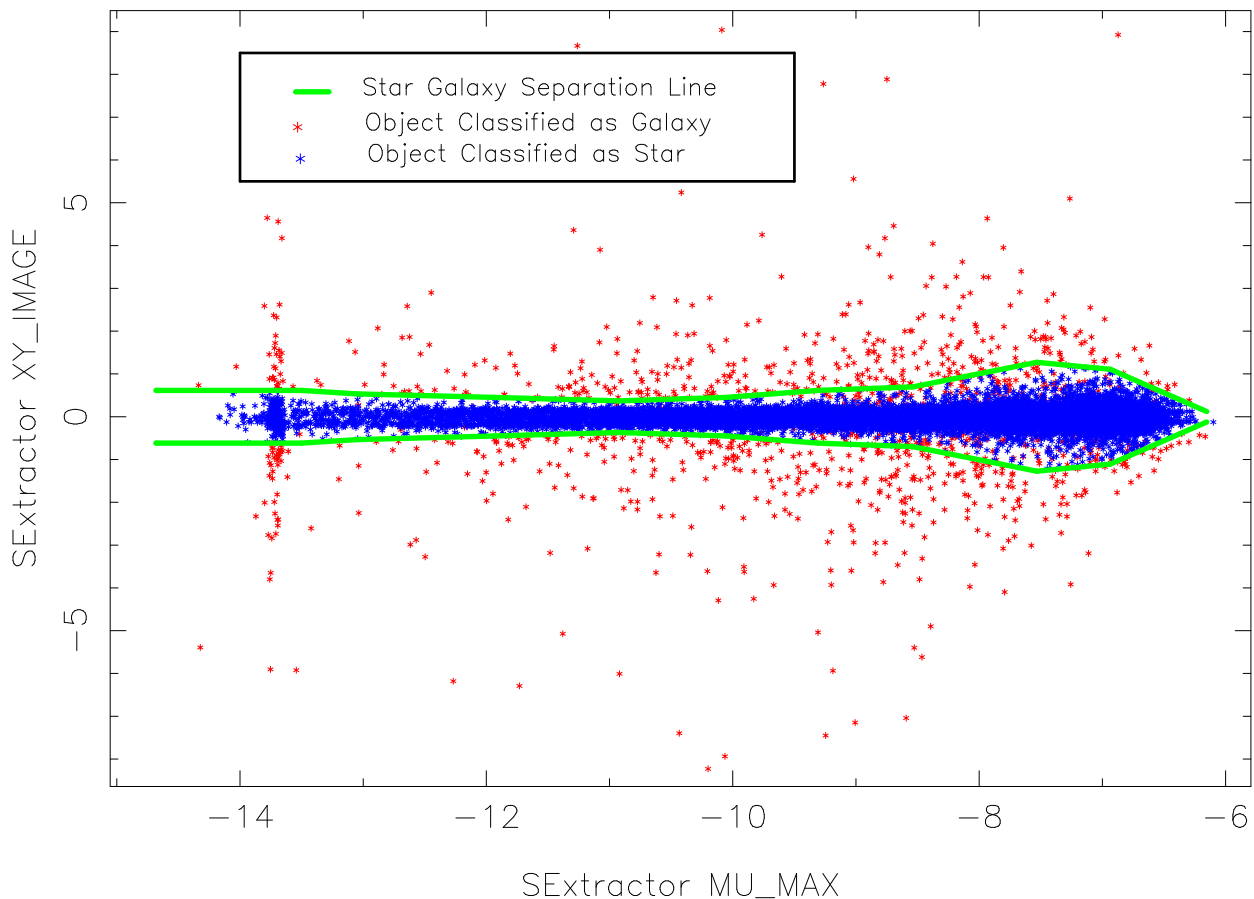,width=\textwidth,angle=-90}
        \caption{Star/galaxy with XY\_IMAGE versus MU\_MAX.  The plot represents all points on amplifier "A" of 90Prime during a 120 second R band exposure  on the night of May 26, 2006.  The discrimination line is plotted in green.  Objects which are classified as a galaxy when all three discriminators are applied are plotted in red while objects which remain on the stellar locus are plotted in blue.  The spike in the XY\_IMAGE direction which occurs at MU\_MAX $\sim$ 13.7 is due to charge bleed for the brightest objects.  Since these objects are already saturated they are not contaminating our star count samples. }

     \label{fig4}
  \end{center}
\end{figure*}

The neural network classifier in Source Extractor does not work on 
images with pixel scales similar to those in our data. We therefore developed
our own simple classifier, called JAL, based on image parameters from Source 
Extractor 
sensitive to the differences between point sources and extended objects. 
We conducted a series of tests of the various image parameters to find three that were suitable  
for a simple parameter space cut which could be applied to each image.  The following three parameter space 
comparisons were adopted:

\begin{enumerate}
\item{\bf{MU\_MAX vs. ISOAREA\_IMAGE }\rm captures the tendency of extended objects to have a surface brightness much lower than a star  at the same isophotal diameter.   A sample parameter space 
 is shown in Figure~\ref{fig2},  with the discriminator curve separating stars from galaxies.}
\item{\bf{MAG\_ISO vs. MAG\_APER }\rm isolates the tendency of extended objects to have a substantial fraction of its light outside of the core of the PSF, generating an isophotal magnitude larger than the aperture magnitude.  An example of this parameter space is shown in Figure~\ref{fig3}, with the discriminator curve separating stars from galaxies.}
\item{\bf{XY\_IMAGE vs. MU\_MAX }\rm shows how an extended object tends to have a moment in the intensity distribution much different than a star at the same peak surface brightness.  The sample parameter space for these parameters is shown in Figure~\ref{fig4},  with the discriminator curve separating stars from galaxies.  This is the weakest of the three criteria and is therefore given one-half the weight  of the other tests.}
\end{enumerate}

We used these three discriminators in a simple  Perl/Pgplot routine 
which allowed the user to interactively define a curve separating stars 
(scored as 0) from galaxies (scored as 1).  The net classification for a given 
image was a weighted score of 0 to 1 averaged from the three tests. During 
the final catalog matching an averaged net score was then computed from the  
V and R bands only.  We conservatively considered an object with a average score greater than 0.3 to be a galaxy.

We developed this method by comparison with other classifiers. 
Training and fine-tuning the parameter spaces was done using stars and 
galaxies classified with  FOCAS in the  Osmer and Hall standard fields 
observed  each night.  As a final test, we chose a 90Prime field (H055+42) 
in the  MAPS Catalog of the POSS I with object classification using a neural network \citep{1992AJ....103..318O,1993PASP..105.1354O} which was  also  observed and classified by SDSS.  The results of our comparison as a function of magnitude are in Tables~\ref{tbl-1} and ~\ref{tbl-2} for the MAPS and SDSS, respectively.  For $V > 16$, our classifier agreed with the APS classification 89\% of the time or better and  with SDSS 95\% of the time or better.  The lower agreement with the MAPS can be attributed to blended images on the MAPS scans which were classified as non-stellar or  galaxies.  The poor agreement with both APS and SDSS for $V < 16$ is easily explained by the onset of saturation in the 90Prime images for our exposure times.

\subsection{Comparison with SDSS}

\begin{figure*}
  \begin{center}
    \leavevmode
      \psfig{file=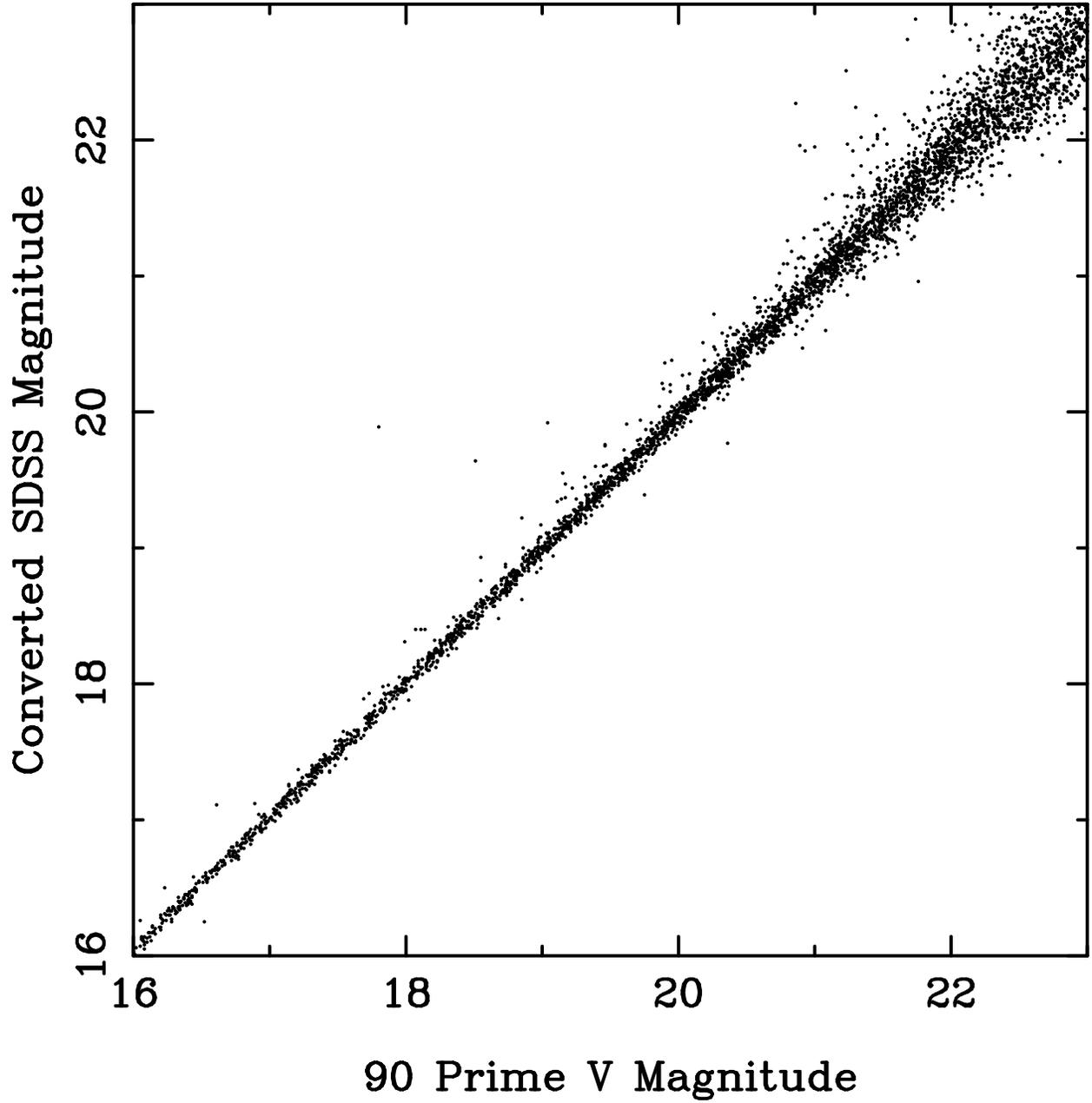,width=\textwidth,angle=-90}
        \caption{Photometric comparison for  H0055+42 between the 90Prime V band data and the SDSS DR5, converted to V band using the transformations of \cite{2006AJ....132..989R}.}
     \label{fig5}
  \end{center}
\end{figure*}

Six of  our program fields overlap SDSS.  Figure~\ref{fig5} shows a photometry comparison between objects classified as stars by both SDSS and our classifier for the program field H055+42.  The SDSS stars have been transformed to the standard V band \citep{2006AJ....132..989R}.  For this comparison we find a zero point difference of 0.02 magnitudes.  For $V \le 20.0$ the RMS difference between SDSS magnitudes and ours is 0.04 magnitudes, for $20.0 < V \le 21.5$ the deviation is 0.09 magnitude and for $V > 21.5$ it is 0.31 magnitude.  
Our data is thus well-suited for our star count analysis down to  $V \sim 21$.

\subsection{Extinction Corrections}

Our magnitudes and colors were corrected for interstellar extinction using the
maps from \cite{sch98}  and  the standard interstellar extinction law \citep{1989ApJ...345..245C}.  Each 
star was corrected based on its coordinates using the interpolation routines
provided.  This approach assumes that the extinction comes
 from relatively near the Sun  and is applied as a constant correction.  
Because our fields are intermediate to high galactic latitudes, extinction is relatively low.  
All but two fields have an average $E(B-V)$ of less than 0.08. One high extinction field 
(H300-20) has an average $E(B-V)$ of 0.36. The extinction-corrected color-magnitude diagram 
resembled the other fields with the placement of the blue ridge at the expected $B-V$ color.  
Therefore extinction anomalies do not play a role in our evaluation of the star counts.

\subsection{Completeness Estimates}

The classical estimate of the completeness limit is determined from a plot of 
the log of the cumulative star counts vs. magnitude and determining 
the magnitude at which it deviates from the  expected straight power law.
For galactic structure work at the magnitudes we are considering however, 
this classical method does not work well because the
density of Thick Disk stars changes with distance above the plane of the 
Galaxy.  
At fainter magnitudes there are simply fewer stars to be found
at larger distances which  can look like incompleteness.  
 
\begin{figure*}
  \begin{center}
    \leavevmode
      \psfig{file=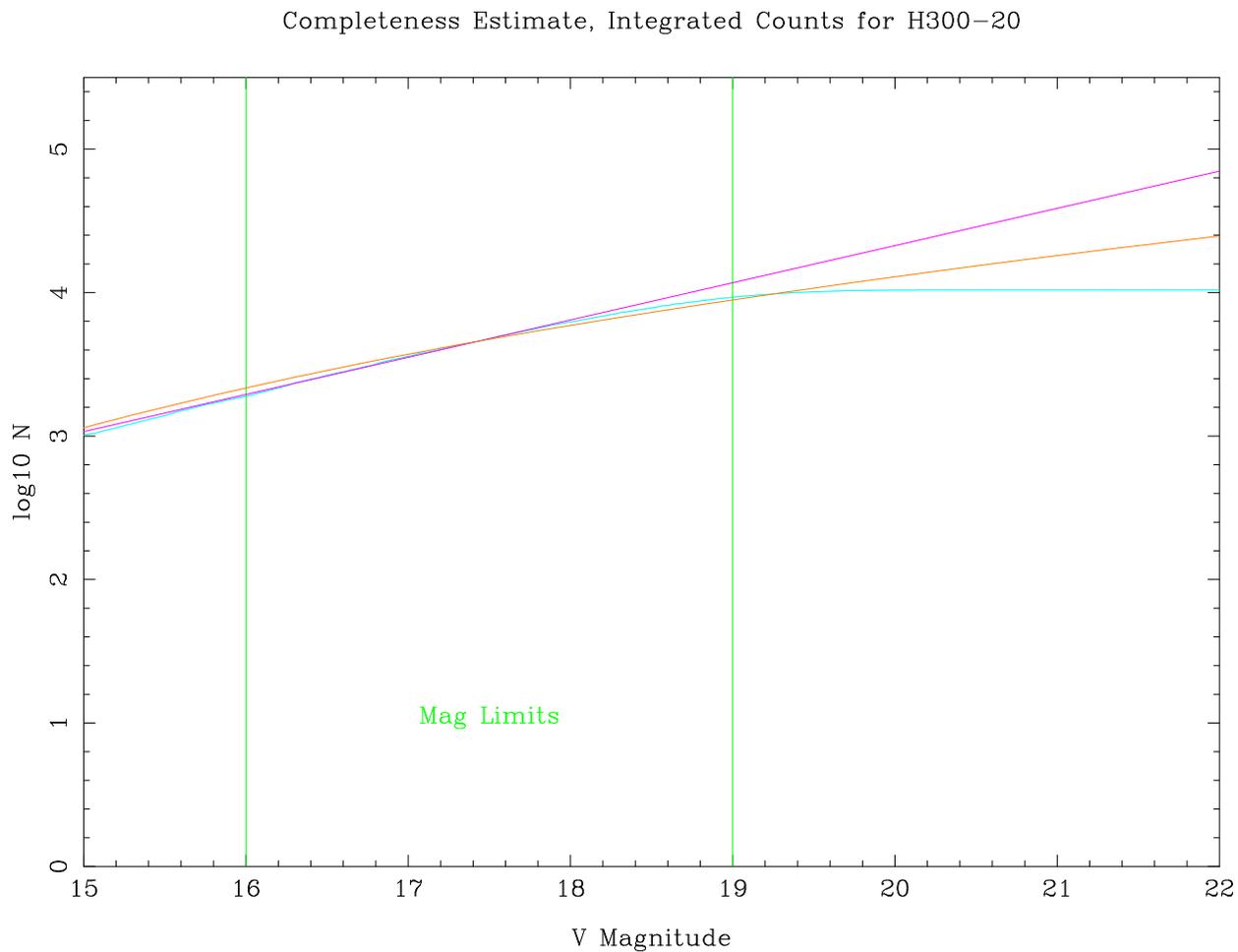,width=\textwidth,angle=-90}
        \caption{Integrated counts for field H300-20 as observed from CTIO (blue curve).  Magnitude limits of interest in this paper ($16 < V < 19$) are delineated with green lines.  Completeness estimates are also plotted:  a classic power law matched to the counts for $V < 19$ (magenta curve) and a model-based estimate using the best fit parameters from \cite{lar03} (orange curve).   The power law implies a completeness limit more than a magnitude brighter the model-based method.}
     \label{fig6}
  \end{center}
\end{figure*}

To derive our completeness estimates we therefore adopt a model-based estimate 
of completeness using Larsen's galactic model program GALMOD \citep{lar03}.  
In Figure~\ref{fig6} we show the results of a completeness estimate for field 
H300-20  using a classical power law compared with one  using the galactic 
model.  While the power law deviates by more than 10\% from the observed 
cumulative counts as bright as $V = 18.2$, the model based estimate shows that 
due to the decreasing  density of stars along the line of sight the completeness limit of the field is closer to $V = 19.6$.  We use the model-based completeness estimate for the rest of our analysis. Our completeness limits are tabulated in Table~\ref{tbl-3}.  If the completeness limit appears fainter than $V = 21.0$, it is reported as being 21.0.  In general, due to their brighter limiting 
magnitudes, our CTIO fields are not as complete as the fields observed
with the Bok telescope.

\section{The Star Counts Ratios -- A Test for Triaxiality}

\begin{figure*}
  \begin{center}
    \leavevmode
      \psfig{file=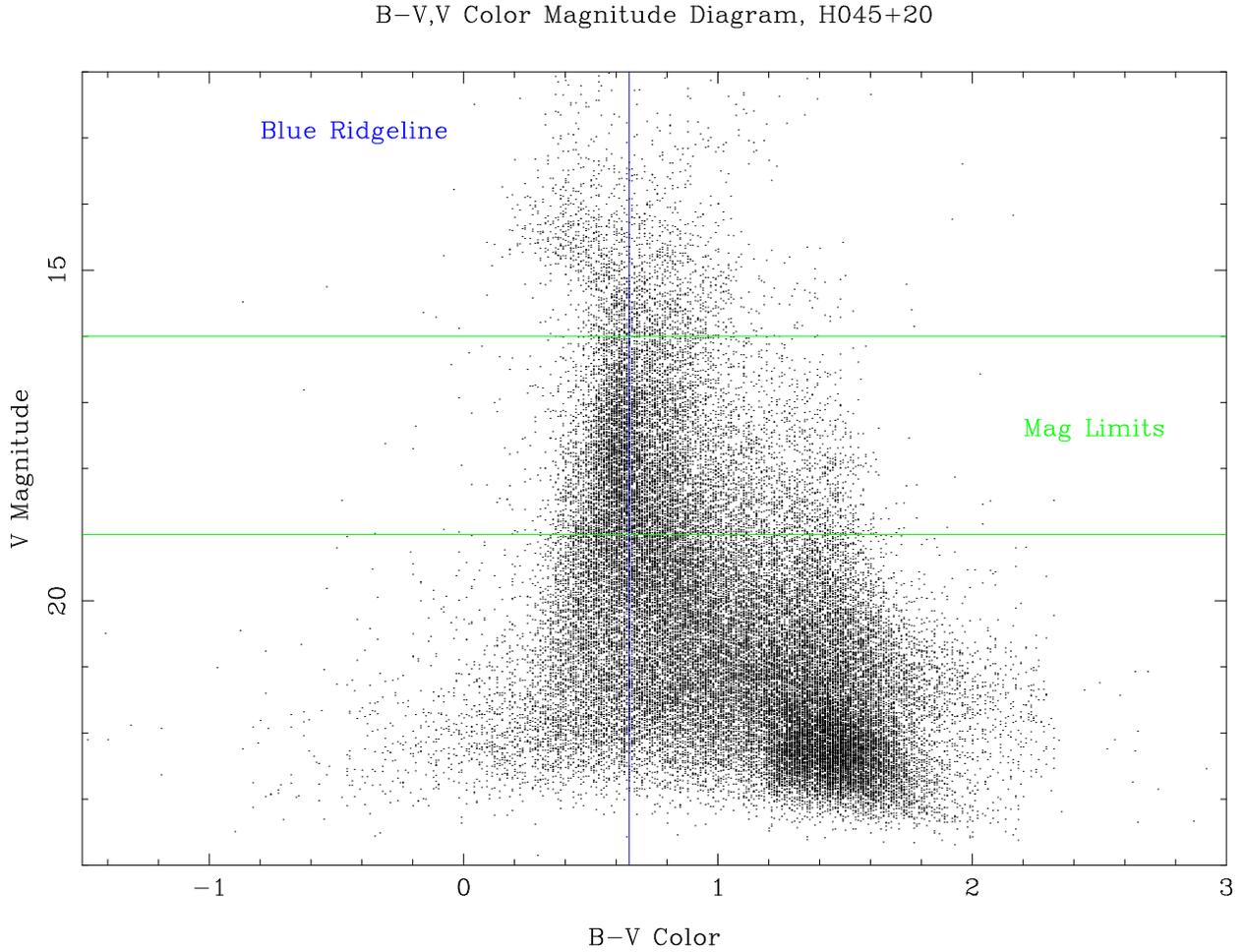,width=\textwidth,angle=-90}
        \caption{Example Color-Magnitude diagram in V and B-V for H045+20, observed using 90Prime.  The magnitude range used in this analysis is bounded in green and the location of the ``blue ridge" is shown in blue.}
     \label{fig7}
  \end{center}
\end{figure*}

The fields used to test for triaxiality are circled 
on Figure~\ref{fig1} and listed in  Table~\ref{tbl-3}.  For this initial 
analysis we use $V$ and the $B-V$ color only. A sample color-magnitude diagram 
for one of our fields is shown in Figure~\ref{fig7}.  

\begin{figure*}
  \begin{center}
    \leavevmode
      \psfig{file=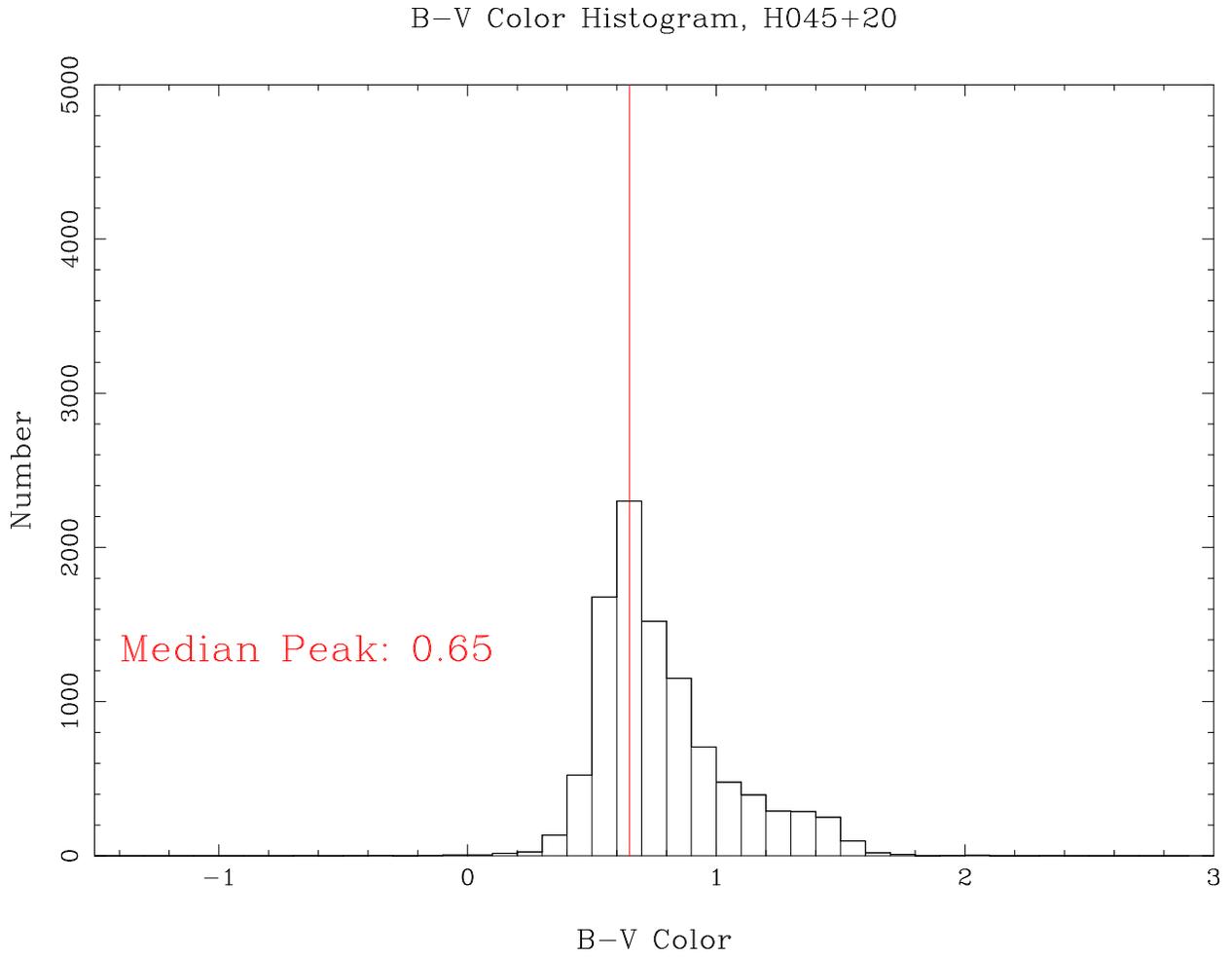,width=\textwidth,angle=-90}
        \caption{The number of stars vs. color for H045+20 for all stars with $ 16 < V < 19$  The location of the blue ridge line as determined by our median routine is indicated.}
     \label{fig8}
  \end{center}
\end{figure*}

To isolate the faint blue stars in the Thick Disk/inner Halo which show the stellar excess, we use the easily recognized  ``blue ridge'' which identifies 
the locus of the main sequence stars in each field.  
The green lines  in Figure~\ref{fig7} delineate this
magnitude range over which all of the Y4KCam and 90Prime fields are  complete.
At these magnitudes, a well-defined peak  in the number of stars is apparent near  
  $B-V \approx 0.6$, the blue ridge,  as illustrated  in the color histogram in Figure ~\ref{fig8}.  
  As we have discussed in
  previous papers \citep{lar96,par03}, the GALMOD model shows that bluewards of this peak,
  we isolate a sample of inner halo and thick disk stars in the  magnitude range  
  $16 < V < 19$.  Assuming that these are main sequence
  subdwarfs, these stars will be between
  1 and 5 kpc from the Sun.

We define  the peak of the blue ridge, $(B-V)_P$,  as the median color of the stars  
 within $\pm$ 0.4 dex of the maximum color bin, see Figure ~\ref{fig8}.   
The resulting blue ridge  peak colors are included in Table~\ref{tbl-3}.  
With the magnitude and color range thus defined,  we
also  give the area of each field, the estimated completeness 
limit $V_C$,  and the number 
of stars bluer than $(B-V)_P$ in the  magnitude ranges $16 < V < 19$, $17 < V < 18$, and 
$18 < V < 19$. 

We then determined the star count ratios for the paired fields in Q1 
and Q4 normalizing for the area of the field. 
The ratios for each magnitude range are summarized in   
Table~\ref{tbl-4} with their error and the significance parameter, {\it s} computed as per
\citet{par03}.  The predicted ratios from  GALMOD are also included.
GALMOD ratios were required for all comparisons which involved fields above and below the plane because the Sun is not in the mid-plane of the disk but instead lies slightly above it \citep{1995AJ....110.2183H}.  Ratios of fields above the plane compared to fields below the plane should be slightly 
less than one due to this geometry and the Sun's location in the Northern Galactic hemisphere.  
Comparisons between lines of sight in the same Galactic hemisphere should have a ratio near unity 
if the stellar distribution in the Galaxy is symmetric.

There are four comparisons in Table~\ref{tbl-4}:   Q1 and Q4 fields above the plane,  Q1 and Q4 fields below the plane,  Q1 fields  above with those  
below the plane,  and  the   Q4 fields  above and below the plane.

We find a statistically significant excess in the ratios for the Q1 to Q4 
fields above the plane for the two fields at the lowest longitudes, {\it l} 
of 45$\arcdeg$ and 50$\arcdeg$ that are an extension of the Hercules Thick
Disk Cloud asymmetry.   The fields at the greater longitudes, {\it l} of
55$\arcdeg$, 60$\arcdeg$, 65$\arcdeg$ and 75$\arcdeg$ show no trend of 
systematic deviations from a ratio of 1.0.  
For Q1 above and below the plane, there is a small  excess above the plane in the 
two lowest longitude fields that does not extend to higher longitudes.  
We do not find evidence for an excess when comparing Q1 to Q4 below the plane or when 
comparing the Q4 fields above and below the plane.  
A few deviations from the expected ratios exist in isolated directions/magnitude bins and 
are of interest, but they do not show a systematic trend.

\section{Summary and Conclusions -- A Search for a Triaxial Thick Disk}

Our star count ratios summarized in Table~\ref{tbl-4} show a small excess of
faint blue stars at {\it l} of 45$\arcdeg$ and 50$\arcdeg$ but 
not at the greater
longitudes including the faintest magnitude intervals. These results support 
 Parker et al's (2003) conclusion that the asymmetry feature in Q1 extends to about
55$\arcdeg$. We thus find {\it no evidence that the Hercules Thick Disk Cloud and the
star count excess is due to a triaxial Thick Disk.}  The ratios at {\it l} of
45$\arcdeg$ and 50$\arcdeg$ return to 1.0 at the faintest magnitude bin ($ 18 < V < 19$), suggesting that we may be seeing through the asymmetry feature; a 
possible clue to its extent along our line of sight in this direction. 
At $ V \sim 19$ however, these faint blue stars will be about 2 kpc above the plane, and may
not be in the Thick Disk.

We also note that the
above vs. below the plane ratios in Q1 and Q4 and the 
Q1/Q4 ratio below the plane are consistent with the predictions of the Galactic model. There is no star count excess. This is not in agreement with \citet{par03}
who concluded that the excess exists both above and below the plane. This may
be evidence that the Hercules Thick Disk Cloud is actually a debris stream.
However, considering the small number of field pairs and the fact that we are apparently at the 
edge of the Cloud at these longitudes, we consider this a tentative result 
pending the analysis of the full set of fields.

Paper II will include an analysis of the full dataset, mapping the stellar excess in Galactic longitude and latitude and along the line of sight to determine
the full extent of the Cloud. Our third paper will include the spectroscopy and analysis of the kinematics of the stars 
showing the excess.

\acknowledgments

This work was supported by  Collaborative National Science Foundation grants to 
Cabanela (AST0729989), Larsen (AST0507309) and Humphreys (AST0507170).
We  thank Steward Observatory and NOAO for observing support, and our respective home institutions for providing facilities support.  Haviland's contributions to this paper were in fulfillment of his Trident Scholarship during his final year at USNA  supported by NRL Grant  N0001409WR40059 (FY09). Cabanela thanks undergraduate research assistants Joshua Swanson and Laura Broaded for testing and reducing the
original Y4KCam data from 2006 April. Larsen also thanks 
Debora Katz for observing assistance at Steward Observatory  and co-advising Haviland's Trident project.

\begin{center}
\begin{deluxetable}{cccccccc}
\tablewidth{0pt}
\tablecaption{Field Information for the Thick Disk Asymmetry Project.\label{tbl-0}}
\tablehead{
\colhead{Field Name} & 
\colhead{$l$} & 
\colhead{$b$} &  
\colhead{Mean RA (J2000)} & 
\colhead{Mean Dec (J2000)} & 
\colhead{Instrument} & 
\colhead{Area} & 
\colhead{Run Observed}  
}
\startdata
H020+20 & 20.63\arcdeg & 18.79\arcdeg & 17$^h$17$^m$12$^s$ & -01\arcdeg46'31'' & 90Prime & 1.04 & 2006 May \\
H020+32 & 20.53\arcdeg & 30.79\arcdeg & 16$^h$35$^m$28$^s$ & +04\arcdeg09'44'' & 90Prime & 1.04 & 2006 May \\
H020+47 & 20.30\arcdeg & 45.70\arcdeg & 15$^h$42$^m$32$^s$ & +11\arcdeg17'18'' & 90Prime & 0.77 & 2008 May \\
H020-47 & 21.02\arcdeg & -48.25\arcdeg & 21$^h$36$^m$27$^s$ & -28\arcdeg00'21'' & Y4KCam & 0.80 & 2008 Oct \\
H023+40 & 23.34\arcdeg & 38.69\arcdeg & 16$^h$12$^m$01$^s$ & +09\arcdeg59'47'' & 90Prime & 0.77 & 2008 May \\
H025+40 & 25.38\arcdeg & 38.83\arcdeg & 16$^h$14$^m$27$^s$ & +11\arcdeg28'33'' & 90Prime & 1.04 & 2006 May \\
H027+37 & 27.31\arcdeg & 35.72\arcdeg & 16$^h$28$^m$36$^s$ & +11\arcdeg28'52'' & 90Prime & 0.77 & 2008 May \\
H027-37 & 28.03\arcdeg & -38.18\arcdeg & 20$^h$59$^m$56$^s$ & -20\arcdeg22'32'' & Y4KCam & 0.80 & 2008 Oct \\
H027+40 & 27.25\arcdeg & 38.73\arcdeg & 16$^h$17$^m$24$^s$ & +12\arcdeg44'26'' & 90Prime & 0.77 & 2008 May \\
H030+20 & 30.51\arcdeg & 18.89\arcdeg & 17$^h$34$^m$43$^s$ & +06\arcdeg29'21'' & 90Prime & 1.04 & 2006 May \\
H030-20 & 30.90\arcdeg & -21.17\arcdeg & 19$^h$58$^m$42$^s$ & -11\arcdeg28'38'' & 90Prime & 0.77 & 2007 Sep \\
H033+40 & 32.68\arcdeg & 38.76\arcdeg & 16$^h$24$^m$25$^s$ & +16\arcdeg37'47'' & 90Prime & 0.77 & 2008 May \\
H033-40 & 34.21\arcdeg & -41.28\arcdeg & 21$^h$19$^m$20$^s$ & -17\arcdeg02'39'' & Y4KCam & 0.48 & 2008 Oct \\
H035+32 & 35.34\arcdeg & 30.89\arcdeg & 16$^h$58$^m$07$^s$ & +15\arcdeg34'43'' & 90Prime & 1.04 & 2006 May \\
H035-32 & 36.06\arcdeg & -33.06\arcdeg & 20$^h$50$^m$41$^s$ & -12\arcdeg23'15'' & 90Prime & 1.04 & 2006 May \\
H042-20 & 43.34\arcdeg & -41.08\arcdeg & 21$^h$30$^m$28$^s$ & -10\arcdeg43'38'' & 90Prime & 0.77 & 2007 Sep \\
H042+40 & 42.09\arcdeg & 38.99\arcdeg & 16$^h$34$^m$08$^s$ & +23\arcdeg35'27'' & 90Prime & 1.04 & 2006 May \\
H042-40 & 43.30\arcdeg & -41.07\arcdeg & 21$^h$30$^m$24$^s$ & -10\arcdeg44'47'' & 90Prime  & 0.77 & 2007 Sep \\
H044+40 & 43.99\arcdeg & 38.92\arcdeg & 16$^h$36$^m$18$^s$ & +24\arcdeg58'46'' & 90Prime & 0.77 & 2008 May \\
H045+20 & 45.43\arcdeg & 19.02\arcdeg & 17$^h$58$^m$57$^s$ & +19\arcdeg18'40'' & 90Prime & 1.04 & 2006 May \\
H045-20 & 45.99\arcdeg & -20.96\arcdeg & 20$^h$24$^m$20$^s$ & +01\arcdeg04'50'' & 90Prime & 1.04 & 2006 May \\
H046+45 & 45.08\arcdeg & 43.79\arcdeg & 16$^h$16$^m$17$^s$ & +27\arcdeg02'01'' & 90Prime & 0.77 & 2008 May \\
H048+45 & 47.27\arcdeg & 43.97\arcdeg & 16$^h$17$^m$05$^s$ & +28\arcdeg36'35'' & 90Prime & 0.77 & 2008 May \\
H050+31 & 50.23\arcdeg & 30.09\arcdeg & 17$^h$20$^m$37$^s$ & +27\arcdeg19'10'' & 90Prime & 1.04 & 2006 May \\
H050-31 & 51.19\arcdeg & -31.87\arcdeg & 21$^h$11$^m$17$^s$ & -00\arcdeg38'31'' & 90Prime & 1.04 & 2006 May \\
H053+42 & 52.79\arcdeg & 41.05\arcdeg & 16$^h$34$^m$02$^s$ & +32\arcdeg02'05'' & 90Prime & 0.77 & 2008 May \\
H055+42 & 54.87\arcdeg & 41.20\arcdeg & 16$^h$34$^m$35$^s$ & +33\arcdeg36'18'' & 90Prime & 1.04 & 2006 May \\
H055-42 & 56.55\arcdeg & -42.86\arcdeg & 21$^h$57$^m$14$^s$ & -03\arcdeg16'54'' & 90Prime & 0.59 & 2007 Sep \\
H060+20 & 60.32\arcdeg & 19.27\arcdeg & 18$^h$21$^m$40$^s$ & +32\arcdeg25'12'' & 90Prime & 1.04 & 2006 May \\
H060-20 & 61.06\arcdeg & -20.69\arcdeg & 20$^h$54$^m$40$^s$ & +12\arcdeg59'39'' & 90Prime & 1.04 & 2006 May \\
H065+31 & 65.08\arcdeg & 30.36\arcdeg & 17$^h$35$^m$16$^s$ & +39\arcdeg46'56'' & 90Prime & 1.04 & 2006 May \\
H065-31 & 66.33\arcdeg & -31.70\arcdeg & 21$^h$41$^m$43$^s$ & +09\arcdeg44'33'' & 90Prime & 0.78 & 2007 Sep \\
H075+20 & 75.28\arcdeg & 19.54\arcdeg & 18$^h$45$^m$52$^s$ & +45\arcdeg44'02'' & 90Prime & 1.04 & 2006 May \\
H075-20 & 76.14\arcdeg & -20.50\arcdeg & 21$^h$32$^m$53$^s$ & +23\arcdeg49'24'' & 90Prime & 0.77 & 2007 Sep \\
H285-20 & 285.37\arcdeg & -20.11\arcdeg & 08$^h$11$^m$31$^s$ & -72\arcdeg10'17'' & Y4KCam & 0.70 & 2008 Oct \\
H285+20 & 286.13\arcdeg & 19.82\arcdeg & 11$^h$23$^m$00$^s$ & -39\arcdeg45'21'' & Y4KCam & 0.90 & 2006 Apr \\
H295-31 & 294.94\arcdeg & -31.35\arcdeg & 04$^h$48$^m$47$^s$ & -82\arcdeg01'58'' & Y4KCam & 0.90 & 2008 Oct \\
H295+31 & 296.39\arcdeg & 30.60\arcdeg & 12$^h$19$^m$37$^s$ & -31\arcdeg22'40'' & Y4KCam & 0.90 & 2006 Apr \\
H300-20 & 300.34\arcdeg & -20.41\arcdeg & 11$^h$28$^m$40$^s$ & -82\arcdeg19'09'' & Y4KCam & 0.85 & 2006 Apr \\
H300+20 & 301.12\arcdeg & 19.50\arcdeg & 12$^h$36$^m$33$^s$ & -42\arcdeg47'15'' & Y4KCam & 0.90 & 2006 Apr \\
H305-42 & 304.75\arcdeg & -42.67\arcdeg & 00$^h$27$^m$26$^s$ & -74\arcdeg56'07'' & Y4KCam & 0.90 & 2008 Oct \\
H305+42 & 306.66\arcdeg & 41.38\arcdeg & 12$^h$58$^m$06$^s$ & -20\arcdeg52'02'' & 90Prime & 1.04 & 2006 May \\
H307+32 & 308.71\arcdeg & 41.23\arcdeg & 13$^h$04$^m$44$^s$ & -20\arcdeg53'57'' & 90Prime & 0.77 & 2008 May \\
H310+31 & 311.27\arcdeg & 30.30\arcdeg & 13$^h$19$^m$48$^s$ & -31\arcdeg29'45'' & Y4KCam & 0.90 & 2006 Apr \\
H312+45 & 314.28\arcdeg & 44.14\arcdeg & 13$^h$20$^m$13$^s$ & -17\arcdeg27'33'' & 90Prime & 0.77 & 2008 May \\
H314+45 & 315.77\arcdeg & 44.09\arcdeg & 13$^h$24$^m$38$^s$ & -17\arcdeg18'20'' & 90Prime & 0.77 & 2008 May \\
H315-20 & 315.40\arcdeg & -20.79\arcdeg & 16$^h$51$^m$22$^s$ & -76\arcdeg49'47'' & Y4KCam & 0.90 & 2006 Apr \\
H315+20 & 316.13\arcdeg & 19.22\arcdeg & 13$^h$52$^m$43$^s$ & -41\arcdeg26'05'' & Y4KCam & 0.90 & 2006 Apr \\
H316+40 & 317.52\arcdeg & 39.04\arcdeg & 13$^h$34$^m$48$^s$ & -21\arcdeg56'24'' & 90Prime & 0.77 & 2008 May \\
H318-40 & 318.02\arcdeg & -41.07\arcdeg & 22$^h$12$^m$05$^s$ & -71\arcdeg50'00'' & Y4KCam & 0.90 & 2008 Oct \\
H318+40 & 319.46\arcdeg & 39.17\arcdeg & 13$^h$40$^m$57$^s$ & -21\arcdeg25'46'' & 90Prime  & 1.04 & 2006 May \\
H325-32 & 325.23\arcdeg & -32.96\arcdeg & 20$^h$10$^m$13$^s$ & -70\arcdeg11'34'' & Y4KCam & 0.90 & 2008 Oct \\
H325+32 & 326.19\arcdeg & 31.05\arcdeg & 14$^h$15$^m$31$^s$ & -27\arcdeg16'07'' & Y4KCam & 0.92 & 2006 Apr \\
H327+40 & 328.88\arcdeg & 38.89\arcdeg & 14$^h$10$^m$41$^s$ & -19\arcdeg12'19'' & 90Prime & 0.77 & 2008 May \\
H330-20 & 330.46\arcdeg & -21.01\arcdeg & 18$^h$10$^m$53$^s$ & -64\arcdeg13'34'' & Y4KCam  & 0.90 & 2006 Apr \\
H330+20 & 330.97\arcdeg & 18.98\arcdeg & 14$^h$59$^m$22$^s$ & -36\arcdeg05'31'' & Y4KCam & 0.90 & 2006 Apr \\
H333-37 & 28.09\arcdeg & -37.78\arcdeg & 20$^h$58$^m$26$^s$ & -20\arcdeg11'59'' & Y4KCam & 0.21 & 2008 Oct \\
H333+37 & 334.20\arcdeg & 35.82\arcdeg & 14$^h$32$^m$34$^s$ & -20\arcdeg03'01'' & 90Prime & 0.77 & 2008 May \\
H333+40 & 334.30\arcdeg & 38.81\arcdeg & 14$^h$26$^m$45$^s$ & -17\arcdeg23'50'' & 90Prime & 0.77 & 2008 May \\
H335-20 & 335.09\arcdeg & -41.24\arcdeg & 21$^h$09$^m$48$^s$ & -60\arcdeg32'57'' & Y4KCam & 0.20 & 2008 Oct \\
H335-40 & 335.07\arcdeg & -41.19\arcdeg & 21$^h$09$^m$26$^s$ & -60\arcdeg34'26'' & Y4KCam & 0.53 & 2008 Oct \\
H335+40 & 336.19\arcdeg & 38.97\arcdeg & 14$^h$31$^m$47$^s$ & -16\arcdeg32'02'' & 90Prime & 1.04 & 2006 May \\
H337+40 & 338.24\arcdeg & 38.76\arcdeg & 14$^h$37$^m$54$^s$ & -15\arcdeg52'59'' & 90Prime & 0.77 & 2008 May \\
H340-20 & 340.52\arcdeg & -21.09\arcdeg & 18$^h$36$^m$21$^s$ & -55\arcdeg23'38'' & Y4KCam  & 0.90 & 2006 Apr \\
H340+20 & 340.86\arcdeg & 18.91\arcdeg & 15$^h$36$^m$16$^s$ & -30\arcdeg46'05'' & Y4KCam & 0.88 & 2006 Apr \\
H340+32 & 341.02\arcdeg & 30.87\arcdeg & 15$^h$04$^m$29$^s$ & -21\arcdeg10'01'' & 90Prime & 1.04 & 2006 May \\
H340+47 & 341.29\arcdeg & 45.75\arcdeg & 14$^h$30$^m$10$^s$ & -08\arcdeg47'15'' & 90Prime & 0.77 & 2008 May \\
\enddata
\end{deluxetable}
\end{center}

\clearpage

\begin{table}
\begin{center}
\caption{Mean extinction and color terms from the 90Prime camera on the Bok 2.3 meter telescope for May of 2006.\label{tbl-5}}
\begin{tabular}{ccc}
\tableline\tableline
Band or Color & Atmospheric Extinction & Color Terms \\
\tableline
$V$ & $0.142 \pm 0.024$ & $-0.032 \pm 0.010$ \\
$BV$ & $0.089 \pm 0.011$ & $1.125 \pm 0.018$ \\
$VR$ & $0.049 \pm 0.008$ & $0.930 \pm 0.010$ \\
$UB$ & $0.232 \pm 0.009$ & $0.957 \pm 0.021$ \\
\tableline
\end{tabular}
\end{center}
\end{table}

\begin{table}
\begin{center}
\caption{Mean extinction and color terms from the Y4KCam camera on the SMARTS 1.0 meter telescope for April of 2006.\label{tbl-6}}
\begin{tabular}{ccc}
\tableline\tableline
Band or Color & Atmospheric Extinction & Color Terms \\
\tableline
$V$  & $0.163 \pm 0.017 $ & $0.096 \pm 0.013 $ \\
$BV$ & $0.123 \pm 0.007 $ & $0.847 \pm 0.033 $ \\
$VR$ & $0.052 \pm 0.019 $ & $0.912 \pm 0.025 $ \\
$UB$ & $0.245 \pm 0.030 $ & $0.877 \pm 0.057 $ \\
\tableline
\end{tabular}
\end{center}
\end{table}

\begin{table}
\begin{center}
\caption{Star-Galaxy Separation Comparison between the JAL classifier and the Automated Plate Scanner for field H055+42. The table presents the number of classified stars (S) and galaxies (G) for each method along with the number of classifications in agreement.  The last columns express the overall agreement and agreement on only the stellar classification between methods. \label{tbl-1}}
\begin{tabular}{cccccccccccc}
\tableline\tableline
B Mag & APS & APS & JAL & JAL & Agree & Disagree & Agree & Disagree & Overall & Stellar \\
Range & S & G & S & G & G & G & S & S & Agreement & Agreement \\
\tableline
  15 - 16 &   4 &   0 &   3 &   1 &   0 &   1 &   3 &   0 &  75\% &  75\% \\
  16 - 17 &  74 &   2 &  73 &   3 &   1 &   2 &  72 &   1 &  96\% &  97\% \\
  17 - 18 & 377 &  33 & 388 &  22 &  10 &  12 & 365 &  23 &  91\% &  96\% \\
  18 - 19 & 635 &  82 & 653 &  64 &  45 &  19 & 616 &  37 &  92\% &  97\% \\
  19 - 20 & 432 & 191 & 499 & 124 & 110 &  14 & 415 &  81 &  84\% &  96\% \\
  20 - 21 & 234 & 241 & 345 & 130 & 110 &  19 & 213 & 131 &  68\% &  91\% \\
  21 - 22 & 137 & 194 & 264 &  67 &  55 &  12 & 123 & 139 &  54\% &  89\% \\
\tableline
\end{tabular}
\end{center}
\end{table}

\begin{table}
\begin{center}
\caption{Star-Galaxy Separation Comparison between the JAL classifier and the Sloan Digital Sky Survey for field H055+42. The table presents the number of classified stars (S) and galaxies (G) for each method along with the number of classifications in agreement.  The last columns express the overall agreement and agreement on only the stellar classification between methods.\label{tbl-2}}
\begin{tabular}{cccccccccccc}
\tableline\tableline
B Mag & SDSS & SDSS & JAL & JAL & Agree & Disagree & Agree & Disagree & Overall & Stellar \\
Range & S & G & S & G & G & G & S & S & Agreement & Agreement \\
\tableline
  15 - 16 &   0 &   4 &   3 &   1 &   1 &   0 &   0 &   3 &  25\% &   0\% \\
  16 - 17 &  63 &  13 &  73 &   3 &   3 &   0 &  63 &  10 &  86\% & 100\% \\
  17 - 18 & 384 &  26 & 388 &  22 &  10 &  12 & 372 &  16 &  93\% &  96\% \\
  18 - 19 & 660 &  57 & 653 &  64 &  41 &  23 & 637 &  16 &  94\% &  96\% \\
  19 - 20 & 498 & 125 & 499 & 124 & 108 &  16 & 482 &  17 &  94\% &  96\% \\
  20 - 21 & 344 & 131 & 345 & 130 & 116 &  14 & 330 &  15 &  93\% &  95\% \\
  21 - 22 & 248 &  83 & 264 &  67 &  64 &   3 & 245 &  19 &  93\% &  98\% \\
\tableline
\end{tabular}
\end{center}
\end{table}

\begin{table}
\begin{center}
\caption{Extinction Corrected Star Counts. \label{tbl-3}}
\begin{tabular}{ccccccc}
\tableline\tableline
Field Name & Area & $V_C$\tablenotemark{a} & $(B-V)_P$\tablenotemark{b} & $16 < V < 19$ & $17 < V < 18$ & $ 18 < V < 19$ \\
\tableline
H045-20 & 1.02 & 21.0 & 0.70 & 3478 & 1241 &  1513 \\
H045+20 & 1.02 & 21.0 & 0.65 & 3944 & 1465  & 1671 \\
H050-31 & 1.02 & 21.0 & 0.64 & 1077 & 368  & 463 \\
H050+31 & 1.02 & 20.7 & 0.68 & 1179 & 423  & 468 \\
H055+42 & 1.02 & 20.5 & 0.61 & 547  & 154  & 260 \\
H060-20 & 1.02 & 21.0 & 0.66 & 2689 & 949  &  521\tablenotemark{c} \\
H060+20 & 1.00 & 21.0 & 0.72 & 2803 & 1013 &  551\tablenotemark{c} \\
H065+31 & 1.01 & 20.5 & 0.70 &  937 & 342  & 354 \\
H065-31 & 0.72 & 21.0 & 0.66 &  741 & 259  & 269 \\
H075+20 & 1.02 & 21.0 & 0.67 & 2092 & 713  & 774 \\
H075-20 & 0.77 & 19.3 & 0.65 & 1889 & 642  & 683 \\
H285+20 & 0.93 & 19.6 & 0.65 & 1889 & 670  & 735 \\
H285-20 & 0.73 & 19.0 & 0.63 & 1738 & 596  & 632 \\
H295+31 & 0.93 & 19.4 & 0.59 & 890  & 299  & 383 \\
H295-31 & 0.93 & 19.2 & 0.69 &  933 & 285  & 452 \\
H300-20 & 0.93 & 18.5 & 0.58 & 2535 & 992  &  494\tablenotemark{c} \\
H300+20 & 0.93 & 19.2 & 0.62 & 2522 & 932  &  529\tablenotemark{c} \\
H305+42 & 1.02 & 21.0 & 0.64 & 525  & 170  & 242 \\
H310+31 & 0.91 & 19.2 & 0.58 & 870  & 289  & 399 \\
H315-20 & 0.93 & 19.4 & 0.58 & 3176 & 1052 &  1536 \\
H315+20 & 0.93 & 19.5 & 0.58 & 2880 & 980  &  1418 \\
\tableline
\end{tabular}
\tablenotetext{a}{Estimated Completeness Limit for the Field in the V Band.}
\tablenotetext{b}{Median Location for the blue ridge Line.}
\tablenotetext{c}{Due to incompleteness of H300-20, last bin is only to V=18.5.}
\end{center}
\end{table}

\begin{deluxetable}{ccccccccccccc}
\tabletypesize{\scriptsize}
\rotate
\tablecaption{Star Count Ratios for fields with $|l| \ge 45 - 75\arcdeg$ from the Galactic Center.\label{tbl-4}}
\tablewidth{0pt}
\tablehead{
Field & $l_1$/$l_2$ & $b_1$/$b_2$ &\multicolumn{3}{c}{GALMOD Ratio Predictions} & \multicolumn{6}{c}{Observed Count Ratios} \\
Ratio & & &$16 < V < 19$ & $17 < V < 18$ & $18 < V < 19$ & $16 < V < 19$ & $s$ & $17 < V < 18$ & $s$ & $18 < V < 19$ & $s$ \\
}
\startdata
\multicolumn{12}{l}{Quadrant 1/Quadrant 4 ratios above the Galactic Plane}\\
\tableline
H045+20/H315+20  & 45/315 & +20/+20 & 1.00 & 1.00 & 1.00 & $1.25 \pm 0.03 $ & 8.12 &  $1.36 \pm 0.06 $ & 6.45 &  $1.07 \pm 0.04 $ & 1.92  \\
H050+31/H310+31  & 50/310 & +31/+31 & 1.00 & 1.00 & 1.00 & $1.21 \pm 0.05 $ & 3.87 &  $1.31 \pm 0.10 $ & 3.07 &  $1.05 \pm 0.07 $ & 0.65  \\
H055+42/H305+42  & 55/305 & +42/+42 & 1.00 & 1.00 & 1.00 & $1.04 \pm 0.06 $ & 0.66 &  $0.91 \pm 0.10 $ & 0.90 &  $1.07 \pm 0.10 $ & 0.70  \\
H060+20/H300+20  & 60/300 & +20/+20 & 1.00 & 1.00 & 1.00 & $1.03 \pm 0.03 $ & 1.19 &  $1.01 \pm 0.05 $ & 0.24 &  $0.97 \pm 0.06 $ & 0.53  \\
H065+31/H295+31  & 65/295 & +31/+31 & 1.00 & 1.00 & 1.00 & $0.97 \pm 0.05 $ & 0.67 &  $1.05 \pm 0.08 $ & 0.64 &  $0.85 \pm 0.06 $ & 2.37  \\
H075+20/H285+20  & 75/285 & +20/+20 & 1.00 & 1.00 & 1.00 & $1.01 \pm 0.03 $ & 0.30 &  $0.97 \pm 0.05 $ & 0.57 &  $0.96 \pm 0.05 $ & 0.81  \\

\multicolumn{12}{l}{Quadrant 1/Quadrant 4 ratios below the Galactic Plane}\\
\tableline
H045-20/H315-20  & 45/315 & -20/-20 & 1.00 & 1.00 & 1.00 & $1.00 \pm 0.02 $ & 0.06 &  $1.08 \pm 0.05 $ & 1.68 &  $0.90 \pm 0.03 $ & 3.13  \\
H060-20/H300-20  & 60/300 & -20/-20 & 1.00 & 1.00 & 1.00 & $0.97 \pm 0.03 $ & 1.23 &  $0.87 \pm 0.04 $ & 3.23 &  $0.96 \pm 0.06 $ & 0.64  \\
H065-31/H295-31  & 65/295 & -31/-31 & 1.00 & 1.00 & 1.00 & $1.03 \pm 0.03 $ & 0.89 &  $1.02 \pm 0.06 $ & 0.37 &  $1.02 \pm 0.06 $ & 0.43  \\
H075-20/H285-20  & 75/285 & -20/-20 & 1.00 & 1.00 & 1.00 & $1.03 \pm 0.05 $ & 0.51 &  $1.17 \pm 0.10 $ & 1.73 &  $0.77 \pm 0.06 $ & 3.91  \\

\multicolumn{12}{l}{Quadrant 1 ratios above/below the Galactic Plane}\\
\tableline
H045+20/H045-20  & 45/45 & +20/-20 & 0.94 & 0.94 & 0.97 & $1.13 \pm 0.03 $ & 6.33 &  $1.18 \pm 0.05 $ & 3.96 &  $1.10 \pm 0.04 $ & 3.25  \\
H050+31/H050-31  & 50/50 & +31/-31 & 0.97 & 0.97 & 0.98 & $1.09 \pm 0.05 $ & 2.40 &  $1.15 \pm 0.08 $ & 2.25 &  $1.01 \pm 0.07 $ & 0.43  \\
H060+20/H060-20  & 60/60 & +20/-20 & 0.94 & 0.93 & 0.96 & $0.99 \pm 0.03 $ & 1.66 &  $1.06 \pm 0.05 $ & 1.80 &  $1.07 \pm 0.07 $ & 1.85  \\
H075+20/H075-20  & 75/75 & +20/-20 & 0.94 & 0.93 & 0.96 & $0.84 \pm 0.05 $ & 2.00 &  $0.84 \pm 0.05 $ & 1.80 &  $0.86 \pm 0.04 $ & 2.50  \\
H065+31/H065-31  & 65/65 & +31/-31 & 0.97 & 0.96 & 0.98 & $0.90 \pm 0.04 $ & 1.75 &  $0.94 \pm 0.08 $ & 0.25 &  $0.94 \pm 0.08 $ & 0.50  \\

\multicolumn{12}{l}{Quadrant 4 ratios above/below the Galactic Plane}\\
\tableline
H315+20/H315-20  & 315/315 & +20/-20 & 0.94 & 0.94 & 0.97 & $0.91 \pm 0.02 $ & 1.50 &  $0.93 \pm 0.04 $ & 0.25 &  $0.92 \pm 0.03 $ & 1.60 & \\
H300+20/H300-20  & 300/300 & +20/-20 & 0.94 & 0.93 & 0.96 & $0.99 \pm 0.03 $ & 1.60 &  $0.94 \pm 0.04 $ & 0.25 &  $1.07 \pm 0.07 $ & 1.56 & \\
H285+20/H285-20  & 285/285 & +20/-20 & 0.94 & 0.93 & 0.96 & $0.85 \pm 0.04 $ & 2.25 &  $0.88 \pm 0.05 $ & 1.00 &  $0.91 \pm 0.05 $ & 1.00 & \\
H295+31/H295-31  & 295/295 & +31/-31 & 0.97 & 0.96 & 0.98 & $0.95 \pm 0.04 $ & 1.03 &  $1.05 \pm 0.09 $ & 1.00 &  $0.85 \pm 0.06 $ & 2.16 & \\
\enddata
\end{deluxetable}

\end{document}